\documentstyle[
cite,
sprocl,
epsfig]{article}%

\def\Rf#1#2#3#4{{#1} {\bf #2}, #3 (19#4)}

\def\PRp{\em Phys. Rep.}
\def\NPA{{\em Nucl. Phys.} A}
\def\NPB{{\em Nucl. Phys.} B}
\def\PL{{\em Phys. Lett.}  B}
\def\IJMP{{\em Int. J. Mod. Phys.} A}

\def\PRL{\em Phys. Rev. Lett.}
\def\PRp{\em Phys. Reports}
\def\PRD{{\em Phys. Rev.} D}
\def\PRC{{\em Phys. Rev.} C}
\def\ZPC{{\em Z. Phys.} C}

\def\ib{\em ibid.}
\def\ea{\em et al.}
\def\wea{E.K. Sarkisyan {\ea},}

\def\ct{\cite}
\def\vs{\vspace}
\def\be{\begin{equation}}
\def\ee{\end{equation}}
\def\bea{\begin{eqnarray}}
\def\eea{\end{eqnarray}}
\def\lb{\label}
\def\bef{\begin{figure}}
\def\enf{\end{figure}}
\def\bi{\bibitem}
\def\fn{\footnote}

\def\rd{\rm d}
\def\rT{\rm T}
\def\e{\eta}

\def\ve{\varepsilon}
\def\vt{\vartheta}
\def\ef{\stackrel{\sim}{\e}}
\def\ecf{\ef_{\rm 0}}
\def\al{\langle}
\def\ar{\rangle}
\def\De{\Delta\e}

\def\dep{\delta$$\ef}
\def\dn{\delta n}
\def\r{\rho}
\def\ddf{\r(\ef)}
\def\mn{\rm min}
\def\mx{\rm max}
\def\se{\simeq}
\def\c2{\chi^2/{\rm DOF}}
\def\st{(\rm stat)}
\def\sy{(\rm syst)}
\def\rmx{\r_{\mx}}
\def\rmxa{\al \rmx \ar }
\def\ma{14$$<$$n$$<$$20}
\def\mb{24$$<$$n$$<$$30}

\def\aln{all $\, n$}

\def\vp{\varphi_q}

\def\la{\lambda_q}

\def\muq{\mu_q}
\def\psq{\psi_{p,q}}

\begin{document}

\title{COHERENCY VS. STOCHASTICITY IN SPIKE PRODUCTION
IN NUCLEAR COLLISIONS AT INTERMEDIATE ENERGIES}

\author{L. K. GELOVANI,$^a$ G. L. GOGIBERIDZE~\fn{On leave from
Institute of Physics, Tbilisi 380077, Georgia.}}

\address{Joint Institute for Nuclear Research, P.O.B. 79\\
Moscow 101000, Russia}

\author{\underline{E. K. SARKISYAN}}

\address{School of Physics and Astronomy, Tel Aviv University,\\
Tel Aviv 69978, Israel}


\maketitle\abstracts{
Multiparticle spike-production process is investigated in central C-Cu
collisions at 4.5 $A$ Gev/c per nucleon.  The study is based on two
different hypotheses - stochastic vs. coherent - of the formation of
spikes. To observe manifestations of the stochastic dynamics,
the non-regularities in the multiplicity distributions are analyzed using
intermittency approach to a possible phase transition as well as the
one-dimensional intermittency model. The
entropy indices are calculated based on the erraticity approach.
Coherency is studied in the framework of the coherent gluon-jet radiation
model. To this end, the spike-center pseudorapidity distributions are
analyzed. Coexistence of the two mechanisms of spike formation process is
discussed.
}

\section{Introduction}
\lb{intro}

The aim of this talk is to compare the results of the studies
\ct{ypl1}$^-$\ct{pl5} of local fluctuations, or spikes, based on two
different approaches to multiparticle production, namely on stochastic
and 
coherent hypotheses. In the framework of the stochastic
approach, the dynamical origin of the fluctuations is ascribed to the
intermittency phenomena, extensively studied in all types of highy-energy
collisions and shown to exist.\ct{rev1} However, despite such an activity,
an origin of the intermittency remains still unclear.
Another possible mechanism of appearance of spikes
could be the coherent particle
emission. Recently such a model,\ct{revd} based on a coherent gluon
radiation picture, has been applied for hadronic collisions.\ct{pp} The
observations have been found to be in agreement with the theoretical
predictions. A study of local fluctuations in coherent vs. chaotic terms
has a specific interest in particle production in nuclear collisions due
to an expectation of quark-gluon plasma formation and its possible
manifestation in stiochastic scenario.\ct{rev1,bnp,ptr} Note that coherent
emission could be a reason of the intermittency effect suppresssion in
nuclear collisions as observed.\ct{rev1}

\section{Data Sample}
\lb{data}

The study is based on a sample of hadrons produced in the interactions of
4.5~$A$ GeV$/c$ carbon, $^{12}$C, nuclei with a copper target inside the 2m
Streamer Chamber SKM-200~\ct{skm1} at the JINR Synchrophasotron
(Dubna).  A central collision trigger was used:  absence of charged
particles with momenta $p>3$ GeV$/c$ in a forward cone of 2.4$^{\circ}$ was
required.  The systematic errors due to the detector effects
were estimated
do not exceed 3\% \ct{skm1}.

The scanning and the handling of the film data were carried out on special
scanning tables of the Lebedev Physical Institute (Moscow).\ct{obr} The
average measurement error in the momentum was 
$\al\ve_p/p\ar$$\simeq $12$\%$, 
and that in the polar angle was
${\al}\ve_{\vt}\ar$$\simeq $$2^{\circ}$. 
The spikes are studied for
charged particles in the pseudorapidity window 
$\De$$=$$0.2$$-$$2.8$
($\e$$=$$-$$\ln\tan (\vt/2)$) 
with the accuracy 
$\al\ve_{\e}\ar$$\stackrel{<}{_\sim}$0.1. In
addition, particles with 
$p_{\rT}$$>$$1$ GeV/$c$ are excluded from the
investigation as far as no negative charged particles were observed with
such a transverse momentum.  Under the assumption of an equal number of
positive and negative pions, this cut was applied to eliminate the
contribution of protons. A total of 663 events has been analyzed with the
average multiplicity of $23.0\pm 0.4$. 

To overcome the effect of the pseudorapidity spectrum shape
and 
to make  the results comparable with other experiments, the
``cumulative''
variable,~\ct{fl}
$$
\ef(\e)\; =
\int_{\e _{\mn}}^{\e}
\r(\e ')\rd \e ' \,
/
\int_{\e _{\mn}}^{\e _{\mx}} \r(\e ')\rd \e '\; ,
$$
with  the uniform spectrum $\ddf$  within the interval [0,1] is
used.

\section{Results}
\lb{res}

\bef[t]	
\begin{flushleft}
\begin{tabular}{lr}
\epsfig{file=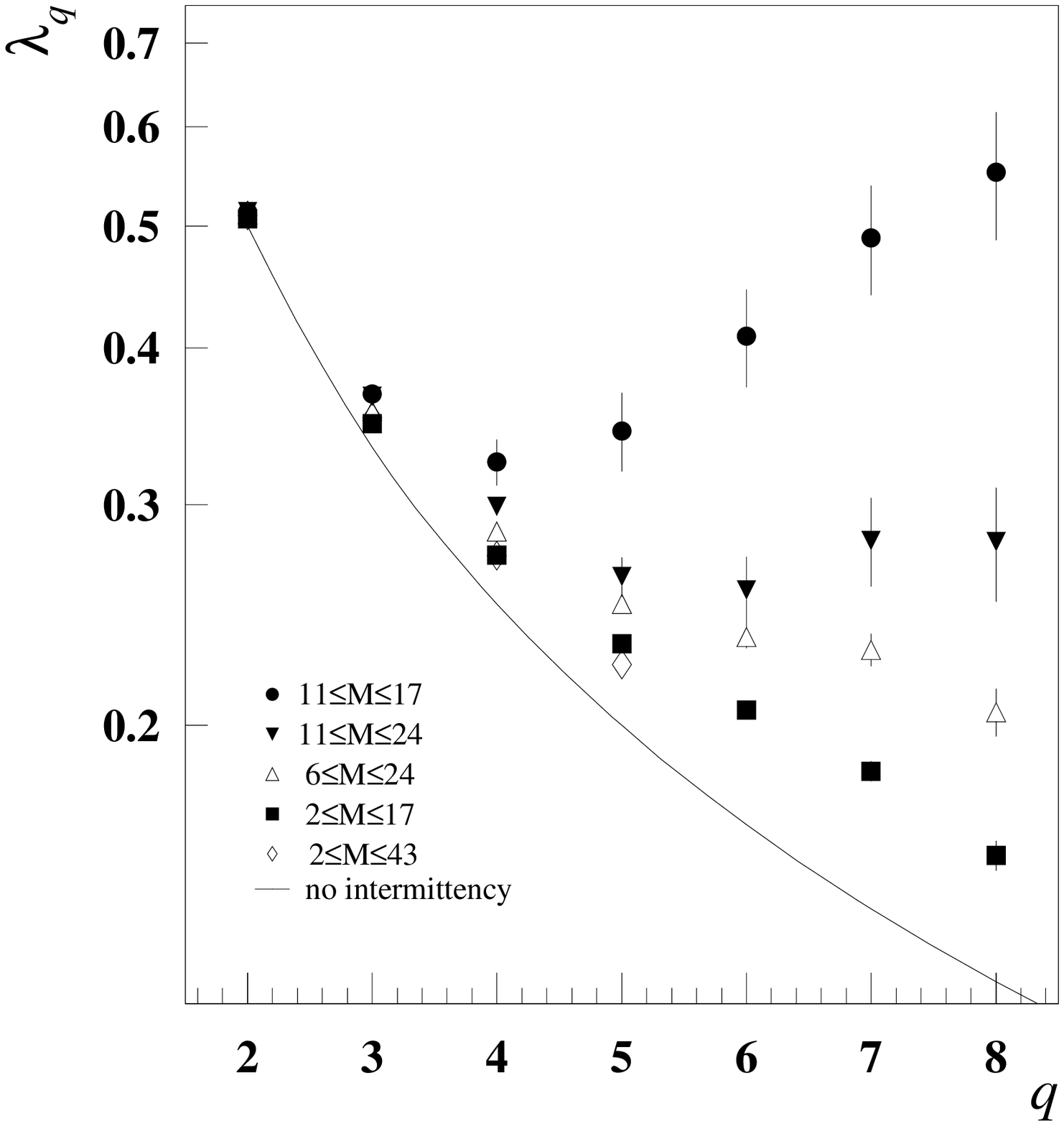,height=2.27in,width=2.2in}
&
\epsfig{file=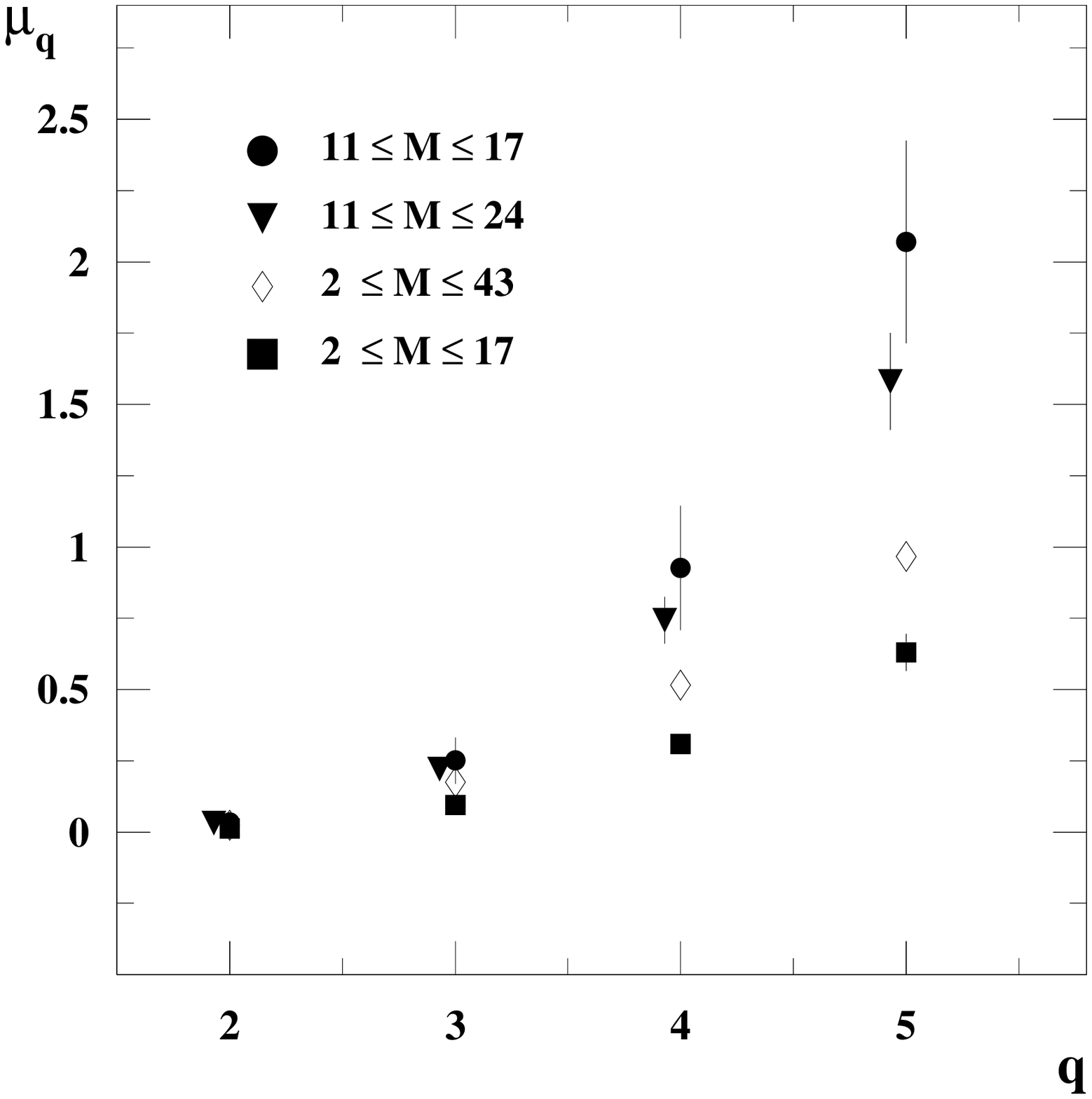,height=2.27in,width=2.05in}
\end{tabular}
\end{flushleft}
\vs{0.17cm}
{\hspace*{1.97cm}{\footnotesize{Figure 1: $\la$ (Eq.~\ref{lam}) vs.  
$q$.~~~~~~~~~~~~~~~~~~~~~~Figure 2: $\muq$
(Eq.~\ref{mue}) vs. $q$.}}
}
\vs{-0.27cm}
\enf

\subsection{Stochasticity Search}
\lb{stoc}

To study stochasticity of the spike-production, we use the
intermittency approach, based on the method of normalised factorial
moments,\ct{bp} and one-dimensional intermittency model.\ct{mxt} 
The intermittency study is extended to search for the fluctuations in the
distributions of the factorial moments that leads to another chaoticity
characteristic such as erraticity.~\ct{erat}

The factorial moments, $F_q$, extracting the $q$-particle
local fluctuations, are predicted to have a power-law increase,
$F_q\propto M^{\varphi_q}$, if the spikes are of a non-statistical
nature. Here, $M$ is number of equal bins into which the 
pseudorapidity subspace is divided.  Such a behavior is called
intermittency and reflects the underlying self-similar dynamics. The
exponents $\varphi_q$, is pointed out to reflect an occurence of possible
phase transition  via the fractal structure of the spike
patterns.\ct{rev1} Monofractality characterizes~\ct{bnp} a second-order
phase transition, while formation of multifractals is assigned~\ct{pesch}
to a self-similar cascading with a possible ``non-thermal''
(non-equilibrium) phase transition.\fn{ To note is that the
thermal phase transition can also lead to multifractality as described in
the framework of the Ginzburg-Landau theory.\ct{ptr} } Multifractality is
found in all types of collisions~\ct{rev1} as
well as in
those studied here.\ct{my3,mg}.

As a signal of the transition, the existence of a minimum of the function
\be 
\la=(\vp+1)/q 
\label{lam} 
\ee 
at a certain ``critical'' value of $q$$=$$q_c$ is expected.\cite{pesch}
However, the minimum of Eq. \ref{lam}  may also be a manifestation of a
coexistence of many small (liquid-type)  fluctuations and a few
high-density ones.\cite{bnp}

Fig. 1 shows the $\la$-function, confirming that at least two
regimes of particle production exist:  one with the phase transition at
4$<$$q_c$$<$5, and another one for which no critical behavior is
reached. The $q_c$-value and the ``critical'' $M$-intervals, which
exhibit
the minimum of $\la$, 
11$\leq$$M$$\leq$17, 11$\leq$$M$$\leq$24, 
are found to
be about the same as in our preceding analyses~\cite{my3} as well as
in recent similar studies in heavy-ion collisions at ultra-high
energies.\cite{uhc}
 
Taking into account the multifractality, the critical $q_c$
indicates a ``non-thermal'' phase transition rather during the
cascade than within one phase. Although the interpretation may be a matter
of debate, it must be noted that the minimum was found earlier also in
hadronic interactions~\ct{rev1} at small $p_{\rT}$ and has been indicated
in high-energy nuclear interactions.\ct{j}

The quantities used in the intermittency approach represent the
averages, for which changes of the density fluctuations from event to
event are not taken
into account. This leads to the loss of information about more structure,
namely, about degree of chaoticity in multiparticle production.  Recently,
erraticity approach has been proposed to take into account the
event-space
(``spatial'') fluctuations.\cite{erat} The method consideres the 
$p$th moment, $C_{p,q}$, of the distributions of the ``horizontal''
normalised factorial moments which have a specific scaling
behavior, $C_{p,q}\propto M^{\psq}$ in the case of self-similarity.
The erraticity indices $\psq$ give a
strength of the ``spatial'' fluctuations.

As a measure of chaoticity in multiparticle production, the entropy
indexes, 
\be
\muq=\frac{\rd }{\rd p}\psq\big|_{p=1}\ ,
\label{mue}
\ee
are considered:  the larger $\mu_q$ is the
more chaotic the system is. 

Fig. 2 shows the $\muq$ calculated~\ct{mg} for different $M$-intervals and
$q=2...5$. The intervals in $M$ are those from Fig. 1, for which different
behavior of the function $\la$ is observed. The large values found for
each interval indicate very chaotic dynamics of particle production,
confirming its cascading nature. It is worthwhile to mention increase
of
the entropy index with approaching to the ``critical'' region.  However,
we must emphasize effect of empty bins at high $q$'s.

To search for dynamical correlations,  we have also used~\ct{ypl1,pl5}
one-dimensional intermittency model~\ct{mxt} that suggests to analyse
maximum density spikes. The key feature of the model is an existence
of two regimes in particle production process - turbulent and
laminar, - leading to two maxima in the maximum density
distributions. Note that the model considers the spikes
selected at given multiplicity $n$ to make an analysis energy and
reaction-type independent and to allow compiling different
experiments results. 

\bef[t]
\begin{center}
\voffset=-2.cm
\epsfig{figure=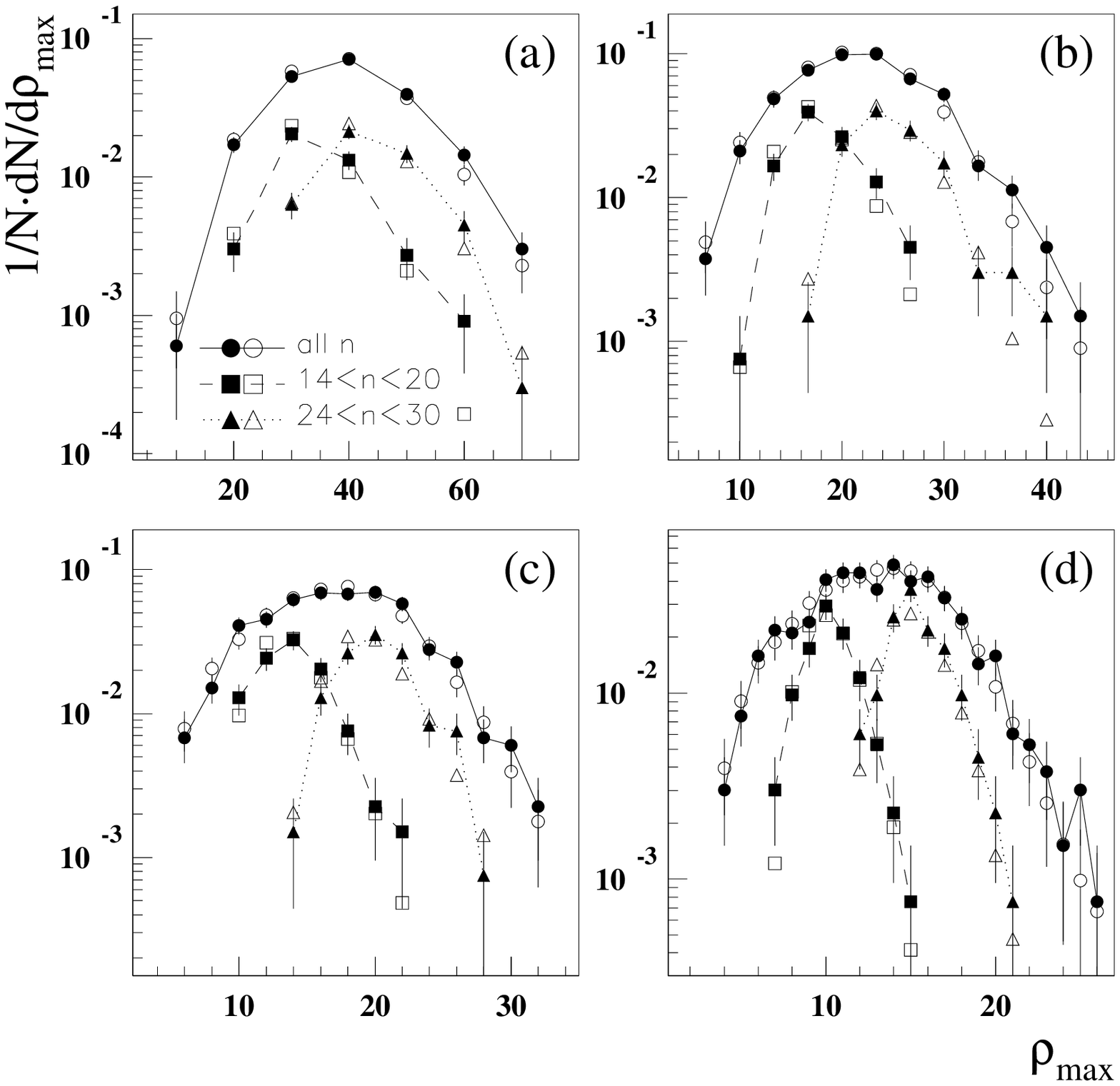,height=3.85in,
width=4.5in}
\end{center}
\vs{.2cm}
\lb{max}
\footnotesize{Figure 3: Normalized experimental (solid
symbols)  and simulated (open symbols) $\rmx$-
distributions
for four $\dep$ and three multiplicity patterns (\aln, $\ma$ and 
$\mb$, respectively):
{\bf (a)} $\dep\, =0.04$, $\c2\se 1.3$, 0.5, 0.7, 
{\bf (b)} $\dep\, =0.12$, $\c2\se 1.2$, 1.3, 2.0, 
{\bf (c)} $\dep\, =0.2$,  $\c2\se 1.1$, 1.0, 1.7,
{\bf (d)} $\dep\, =0.4$,  $\c2\se 0.9$,  0.7, 0.9.        
}
\vs{-.4cm}
\enf

Increased statistics, we have updated~\ct{pl5} our earlier
results~\ct{ypl1} carrying out analysis for different narrow
$n$-intervals as shown in Fig. 3 for two of them along with the
distributions for all $n$. Here, maximum density spike $\rmx$ is defined
as $\dn_{\mx}/\dep$, where $\dn_{\mx}$ is the maximum number of particles
in each event hit in the bin $\dep$. One can see that for the fixed-$n$
intervals the shape of the distributions develops tails at
$\rmx$$>$$\rmxa$, as expected from the model and has indeed been observed
in hadronic interactions.\ct{pp1} The non-poissonian character of the
distributions, expressed as inequality between the dispersion and the
mean values $\rmxa$ (see \ct{pl5}), points at a significant contribution
of the {\it multi}-particle correlations, non-reduceable to the
two-particle ones.\ct{dmx}

To reveal the dynamical correlation effect, the obtained distributions are
compared to those based on the sample of randomly simulated events. The
total of 66300 events were generated, representing independent particle
emission. The
resulting distributions are shown as open symbols in Fig. 3. The values
of $\c2$ tell us that the dynamical correlations are too suppressed by
statistical ``noise'' in these distributions.

A study of the influence of the error $\al\ve_{\vt}\ar$ in the measurement
of the polar angle $\vt$ of the produced charged particles demonstrated
stability of the obtained distributions and, therefore, the reliability of
the conclusions done. 
  
\bef[t]
\begin{center}
\voffset=-2.cm
\epsfig{figure=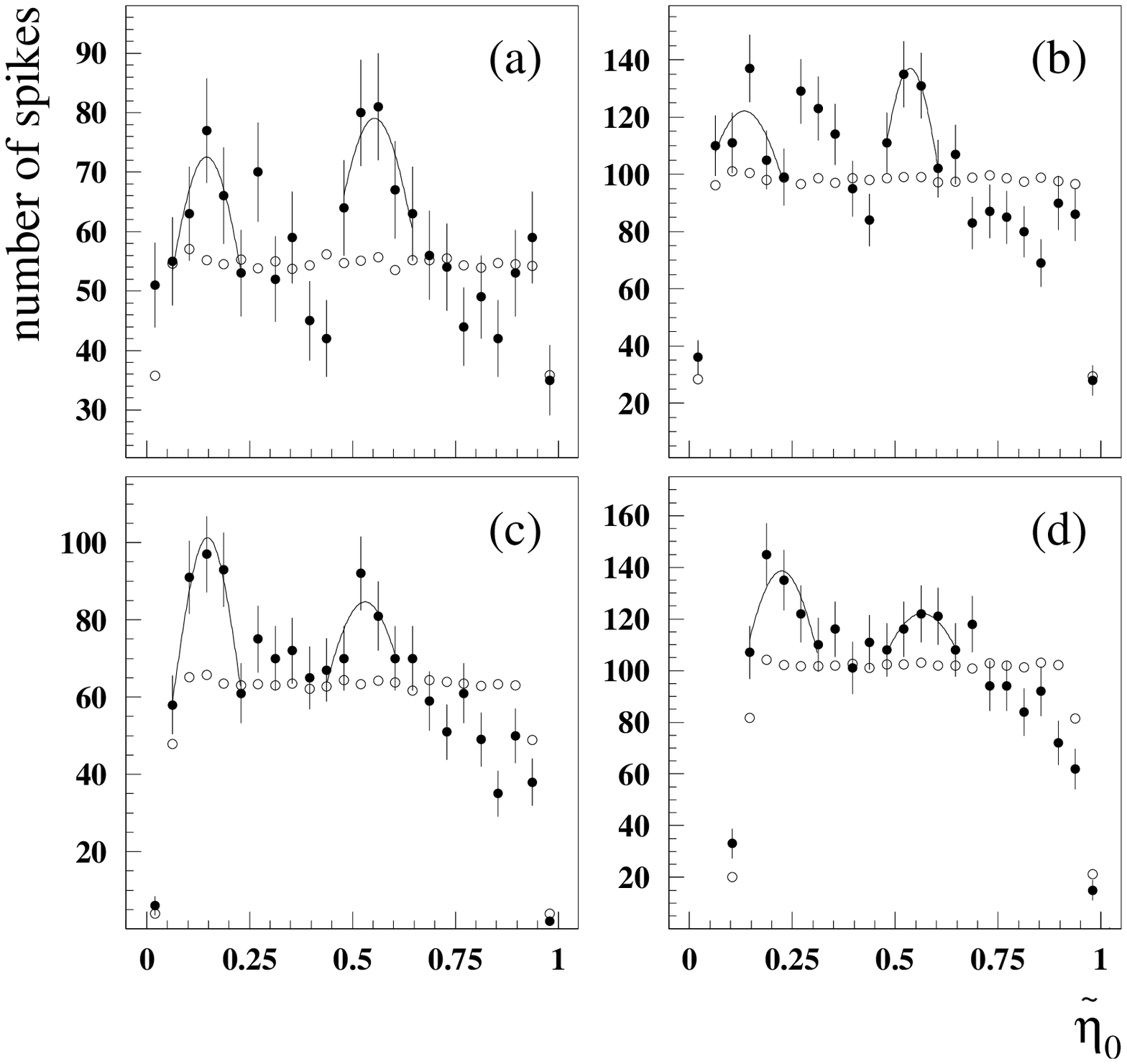,height=3.85in,
width=4.5in}
\end{center}
\vs{.2cm}
\footnotesize{Figure 4: 
Experimental ($\bullet$) and simulated ($\circ$) 
$\ecf$-distributions for  different $\dep$ and various $\dn$: 
{\bf (a)} $\dep\, = 0.04,\; \dn=4$,
{\bf (b)}  $\dep\, = 0.08,\; \dn=5$,
{\bf (c)} $\dep\, = 0.12,\; \dn=7$,
{\bf (d)} $\dep\, =  0.2,\; \dn=9$.
}
\vs{-.6cm}
\lb{centr}
\enf

\subsection{Coherency Search}
\lb{coh}

In the coherency approach~\ct{revd} it is suggested to study the
pseudorapidity spike-center distributions. Acording to the model of
coherent gluon-jet emission, these distributions must have two peaks in
quark-quark radiation (pp collisions) vs. a single peak in
quark-antiquark case (p$\bar{\rm p}$, $\pi$/Kp interactions).  These
structures has recently been observed in hadronic collisions.\ct{pp}

The center of spike, $\ecf$, is determined as
$\ecf=(1/\dn)\sum_{j=1}^{\dn}\ef_j$, where $\dn$ is number of tracks in
spike in each event. Fig. 4 represents the pseudorapidity 
$\ecf$-distributions for different size $\dep$ and for spikes of different
density. Although multi-peak structure can be seen for small $\dep$, two
peaks are well pronounced for larger bins. Fitting these two bumps with
Gaussians and averaging over the different spikes, the peaks are found to
be
placed at $0.17$ and $0.57$. Recounted to the $\e$-variable, the positions
of the peaks are centered at $0.60\pm0.05\st\pm0.12\sy$ and
$1.30\pm0.03\st\pm0.10\sy$ with the distance,
$$
d_0=0.68\pm0.06\st\pm0.16\sy
$$
between them. This value is similar to that found for pp
collisions~\ct{pp}, while the double-peak shape is in agreement with the
predictions of the coherent gluon emission model. Note that $d_0$ is 
higher than that in pp-interactions due to intranuclear processes.

To isolate dynamical correlation effects in these distributions,
analogous distributions have been obtained from the above described
statistical sample of generated events. The $\ecf$-distributions of the
simulated events are shown in Fig.  4 by open circles. One can observe a
remarkable difference between these distributions and those obtained from
data. No any peaks are seen in the latter case, following the background
level and consequently manifesting the double-peak structure in the data
to be a significant one. 

Similarly to the above studies of $\rmx$-spectra, the result was checked
by varying the $\De$-range and the polar angle $\vt$. The  character of
the distributions remains unchanged. 

To note is that a similar structure is observed now in large sample of
Mg-Mg interactions at 4.2 $A$ GeV/c, but for negative pions
only.~\ct{gogf}

\section{Conclusions}

In summary, a study of spike production in central C-Cu collisions at 4.5
GeV/$c$ per nucleon is presented. The analysis considers two different
approaches - stochastic vs. coherent - to the mechanism of spike
formation. Stochasticity search is carried out based on the scaling
properties of the normalised factorial moments and their distributions as
well as using the one-dimensional intermittency model and its prediction
for maximum density distributions. Coherency is serched as it is
described by the model of \v{C}erenkov radiation of gluons at finite
length, predicting specific shapes of spike-center distributions.

In the stochasticity study, multifractality of spike-prodction is
observed, indicating a possible non-thermal phase transition and two
regimes during the cascading. The erraticity approach is used to
calculate the entropy indexes, which points at a chaotic
nature of particle emission process, particularly in a case of the phase
transition. Analysis of maximum density fluctuations show their 
non-poissonian cahracter, indicating a contribution of multiparticle 
correlations to the spikes. 

In studying the spike-center distributions, a double-peak shape is
observed in agreement with the expectation of the coherent gluon
emission model. The distance between the peaks is 
with that found in pp-collisions.

To conclude, a direct study of coherency and stochasticity in spike
appearance in central nuclear collisions at intermediate energy is
performed. Coexistence of these two mechanisms is shown to exist. 

\section*{Acknowledgments} 

E.S. is thankful to the Organizing Committee for inviting him to
participate in this very fruitful Workshop. Thanks go also to our
colleagues  from the GIBS (SKM-200) Collaboration for providing us with
the film data.

\end{document}